\documentclass[aps,twocolumn,preprintnumbers,%
superscriptaddress,floatfix,showpacs]{revtex4}
\usepackage[dvips]{graphics}
\usepackage{amsmath,amsfonts,bm}
\usepackage{epsfig}

\newcommand{\be}{\begin{equation}}
\newcommand{\ee}{\end{equation}}
\newcommand{\bear}{\begin{eqnarray}}
\newcommand{\eear}{\end{eqnarray}}
\newcommand{\ba}{\begin{array}}
\newcommand{\ea}{\end{array}}

\newcommand{\tr}{{\rm tr}}

\newcommand{\bsubeq}{\begin{subequations}}
\newcommand{\esubeq}{\end{subequations}}

\def\bi{\bibitem}

\begin{document}
\preprint{\parbox[b]{1in}{\hbox{\tt
KIAS-P07004 } \hbox{\tt PNUTP-07/A01}\hbox{\tt  hep-th/0701276} }}

\title{Chiral Dynamics of Baryons from String Theory}

\author{Deog Ki Hong}
\email[E-mail: ]{dkhong@pusan.ac.kr} \affiliation{Department of
Physics, Pusan National University,
             Busan 609-735, Korea}
\author{Mannque Rho}
\email[E-mail: ]{mannque.rho@cea.fr} \affiliation{Service de
Physique Th\'eorique, CEA Saclay, 91191 Gif-sur-Yvette, France}

\author{Ho-Ung Yee}
\email[E-mail: ]{ho-ung.yee@kias.re.kr} 
\affiliation{School of Physics, Korea Institute for Advanced
Study, Seoul 130-012, Korea}

\author{Piljin Yi}
\email[E-mail: ]{piljin@kias.re.kr} \affiliation{School of
Physics, Korea Institute for Advanced Study, Seoul 130-012, Korea}

\vspace{0.1in}


\begin{abstract}
We study baryons in an AdS/CFT model of QCD by Sakai and Sugimoto,
realized as  small instantons with fundamental string hairs. We
introduce an effective field theory of the baryons in the
five-dimensional setting, and show that the instanton interpretation
implies a particular magnetic coupling. Dimensional reduction to
four dimensions reproduces the usual chiral effective action, and in
particular we estimate the axial coupling $g_A$ between baryons and
pions and the magnetic dipole moments, both of which are
proportional to $N_c$. We extrapolate to finite $N_c$ and discuss
subleading corrections.
\end{abstract}

\maketitle

\newpage

{\it Introduction}:---
Understanding baryons from the microscopic theory
is a long standing problem, 
since it amounts to solving the low energy QCD, which is strongly
coupled and highly nonlinear. Though limited successes were made
such as in lattice QCD, it is far from complete. Recent discovery of
string/gauge duality~\cite{Maldacena:1997re}, however, enables us to
address the problem, based on holographic models. One interesting
and realistic model among them is the one by Sakai and
Sugimoto~\cite{Sakai:2004cn} (SS model for short). The model
astutely implements chiral symmetry spontaneously broken, and
describes the low-energy dynamics in a manner
consistent with the hidden local symmetry (HLS) theory of the form
developed some years ago with the $\rho$ meson~\cite{HLS}.

In this letter, using the fully five-dimensional picture of baryons which
naturally incorporates the infinite tower of vectors in construction
of the baryon, we will show that chiral dynamics arises naturally in
the large 't Hooft coupling limit
$\lambda=g^2_{YM}N_c\rightarrow\infty$. It has been recognized since
some time that the lowest-lying vector mesons as hidden local fields
could play an important role in the soliton structure~\cite{brihayetal}
and dynamics~\cite{rho} of baryons, which was also recently reconsidered
in the context of the SS model~\cite{nawa}.
Here we find that not just the lowest members but the
{\it whole tower} of the vector fields participate intricately in
the dynamics of baryons, in such a manner that this effectively
simplifies and also relates some four-dimensional interactions.

We start with a brief review of the SS model. In the model, the
stack of $D4$ branes which carries the $SU(N_c)$ pure Yang-Mills theory is
replaced by the dual geometry, when $\lambda\gg1$, with the metric
\begin{equation}
G=\left(\frac{U}{R}\right)^{3/2}\left(ds_4^2+f\,d\tau^2
+\frac{R^3}{U^3}\frac{dU^2}{f}+{R^3\over U} d\Omega_4^2\right)
\end{equation}
where $f(U)=1-U_{KK}^3/U^3$.
The coordinate $\tau$ is periodic with
the period $\delta\tau= {2\pi/ M_{KK}}=4\pi R^{3/2}/3U_{KK}^{1/2}$,
which defines the Kaluza-Klein (KK) mode scale $M_{KK}$ and sets the scale for
massive vector mesons. Note that the parameters of dual QCD are mapped
to the dimensionful parameters here as $R^3={\lambda l_s^2}/{2M_{KK}}$ and
$ U_{KK}={2\lambda M_{KK}l_s^2/9}$ with the string length scale $l_s$.
The string coupling is related
to $SU(N_c)$ Yang-Mills coupling as $ 2\pi g_s={g_{YM}^2}/{M_{KK}l_s}$.

The $D8$ branes, which share coordinates $x^{0,1,2,3}$ with $D4$
branes, are treated as probes and carry $U(N_F)$ Yang-Mills
multiplets from D8-D8 open strings. The induced metric on $D8$ is
\begin{equation} g_{8+1}=g_{4+1}+{R^{3/2} U^{1/2}}d\Omega_4^2\,.
\end{equation}
where the five-dimensional part is conformally equivalent to $R^{3+1}\times I$,
\begin{equation}
g_{4+1}=H(w)\left(dw^2+\eta_{\mu\nu}dx^{\mu}dx^{\nu}\right),
\end{equation}
with $w=\int dU{R^{3/2}}/{\sqrt{U^3-U_{KK}^3}}$ and
$H=(U/R)^{3/2}$. This fifth coordinate is of finite range
$[-w_{max},w_{max}]$ with $w_{max}\simeq {3.64}/{M_{KK}}$. Near
origin $w=0$, we have the approximate relation, $U^3\simeq
U_{KK}^3(1+M_{KK}^2w^2)$.

The main point of this model is that the D8 comes with two
asymptotic regions (corresponding to UV) at $w\rightarrow\pm
w_{max}$, where $U(N_f)$ gauge symmetry of D8 can be each
interpreted as $U(N_f)_{L,R}$ chiral symmetry, respectively,
of fermions from D4-D8 strings.

As D4's are replaced by the
geometry, these D4-D8 strings are connected and become
D8-D8 strings. Thus, mode-expanding the $U(N_F)$ gauge fields on D8
along the fifth direction produces $SU(N_c)$ gauge singlets
which are the pions and the infinite tower
of massive vector mesons. Writing the chiral field as
$\xi(x)=e^{i\pi(x) /f_\pi}$, we have 
in the $A_w=0$ gauge 
\begin{equation}\label{mode}
A(x;w)= i\alpha(x)\psi_0(w)+i\beta(x) +\sum_n
a^{(n)}(x)\psi_{(n)}(w)
\end{equation}
with $ \alpha(x)\equiv \{\xi^{-1},d\xi\}$ and
$\beta(x)\equiv\frac{1}{2}[\xi^{-1},d\xi]$. The zero mode
$\psi_0$
approaches $\pm 1/2$ at the two boundaries. Keeping the pions only
results in the Skyrme Lagrangian for $U=\xi^2$, while including
massive vectors as well produces a theory 
with hidden local symmetries~\cite{HLS}.

{\it Baryons as small and hairy instantons}:---
 A baryon in this model
corresponds to a $D4$ brane wrapping the compact
$S^4$~\cite{witten-baryon}, which is dissolved into D8 as an
$U(N_F)$ instanton. Relation between this and the usual Skyrmion
picture was clarified in~\cite{Sakai:2004cn,Son:2003et}.

In the present curved geometry, with $N_c$ flux in the
background geometry, the compact D4 admits fundamental string
tadpoles which have to be cancelled by $N_c$ fundamental strings
attached to it. The other endpoints of the strings can only go to D8
and thus will behave as electric charge with respect to the trace
part of $U(N_F)$. Thus, a D4 brane on $S^4$ tends to be pulled into
D8 and becomes a finite-size instanton. On the other hand, the background
geometry induces a position-dependent electric coupling for $U(N_F)$
gauge fields, which favors point-like instanton at $w=0$.
The competition between the two will determine the size of the instanton.

The $4+1$ dimensional effective action of $U(N_F)$ Yang-Mills fields
in the conformal coordinate system is
\begin{equation}
\frac14\;\int d^4x dw \;\frac{8\pi^2 R^3U(w)}{3 (2\pi l_s)^5 (2\pi
g_s)} \;\tr F_{mn}F^{mn}
\end{equation}
from which we find the effective electric coupling
\begin{equation}
\frac{1}{e^2(w)}\equiv \frac{8\pi^2 R^3U_{KK}}{3 (2\pi l_s)^5(2\pi g_s)}\frac{U(w)}{U_{KK}}
=\frac{\lambda N_c M_{KK}}{108\pi^3}\frac{U(w)}{U_{KK}}\, .
\end{equation}
A point-like instanton that is localized at $w=0$ would have
the mass $m_B^{(0)}\equiv {4\pi^2}/{e^2(0)}=({\lambda N_c}/{27\pi})M_{KK}$
which is also the mass of D4 wrapping $S^4$ at $w=0$.

If the instanton gets bigger, on the other hand, the configuration
costs more energy, since $1/e^2(w)$ is an increasing function of
$|w|$. For a very small instanton of size $\rho$, this additive
correction to the instanton mass is found to be
$\simeq  m_B^{(0)}M_{KK}^2\rho^2/6$,
using the spread of the instanton density $D(x^i,w)\sim
\rho^4/(x^2+w^2+\rho^2)^4$~\cite{longer}. The competing effects come from the
energy cost of the electric charges, which
arises due to a Chern-Simons
term. One finds the electric charge density is
proportional to $D(x^i,w)$, and the five dimensional Coulomb energy is
readily estimated as \cite{longer}
\begin{equation}
\simeq \frac{e(0)^2N_c^2}{20\pi^2\rho^2},
\end{equation}
provided that $\rho M_{KK} \ll 1$.

The size of the instanton localized at $w=0$ is then determined by
minimizing the sum. This gives the size of the baryon 
\begin{equation}
\rho_{baryon}\simeq \frac{(2\cdot 3^7\cdot\pi^2/5)^{1/4}}{M_{KK}\sqrt{\lambda}}
\sim \frac{9.6}{M_{KK}\sqrt{\lambda }}. \label{size}
\end{equation}
For a large 't Hooft coupling $\lambda$, the baryon size is then
significantly smaller than the relevant scale of the dual QCD, and
the mass correction to the baryon due to its 5-dimensional electric
coupling is also suppressed by $1/\lambda$ compared to $m_B^{(0)}$.
In making this estimate,  we are ignoring the backreaction of the
instanton to the geometry and the position-dependent coupling from
the origin. This probably results in a slight underestimate of
$\rho_{baryon}$, which is controlled by the inverse power of $\lambda$.

{\it Five-dimensional effective action of baryons}:--- Our
first task is to understand the effective action of the instanton
soliton in five dimensions. The soliton by itself is a classical
object. In order to treat it quantum mechanically, one first needs
to quantize their collective coordinates and classify the resulting
particles according to their spin content and the representations
under other symmetries.

 More subtleties come about because the
instanton here is endowed with electric charges, which are remnant
of fundamental strings attached to D4 on $S^4$. Having in mind an
extrapolation to the real QCD, we restrict ourselves to the case
of fermionic baryons with fundamental
representation under $U(N_F=2)$, denoted 
as $\cal B$.

After a suitable
rescaling of the $\cal B$ field in the conformally flat coordinates $(x^\mu,w)$,
we have
\begin{equation}\label{5dfermion}
-i\int d^4 x\,dw\, \left[\bar{\cal B}\gamma^\mu
D_\mu{\cal B}+\bar {\cal B}\gamma^5\partial_w {\cal B} +
m_b(w)\bar{\cal B}{\cal B}\right]\, ,
\end{equation}
in the $A_w=0$ gauge with $D_\mu=\partial_\mu-i A_\mu$ and $(-,+,+,+,+)$ convention.
The gauge field $A$ here is that of $U(N_F)$ on D8, as before,
which encodes the pions and  the entire tower of  massive vector mesons.
$m(w)$ reflects the fact that the instanton costs more energy
if it moves away from $w=0$. This is the minimal set of terms
consistent with the diffeomorphism and gauge invariance. 

However, this cannot be the complete form of the baryon action at low energy.
Since the baryon is represented by a small instanton
soliton with a long-range tail of self-dual gauge field $F\sim
\rho_{baryon}^2/r^4$,
there should be a coupling between
a ${\cal B}$ bilinear and the five-dimensional gauge field such that
each ${\cal B}$-particle generates the  tail on $F$.
There is a unique vertex that does the job,
\begin{eqnarray}
\int d^4 x dw\left[g_5(w){\rho_{baryon}^2\over e^2}\bar{\cal
B}\gamma^{mn}F_{mn}{\cal B} \right]\,.\label{magnetic}
\end{eqnarray}
Writing the upper 2-component part of ${\cal B}$ as ${\cal U}\,e^{-iEt}$,
and approximating $m_b$ by its central value, we find the on-shell condition
is solved by
\begin{equation}
{\cal B}=\left(\begin{array}{r}{\cal U}\\ \pm i {\cal U}\end{array}
\right)e^{\mp im_b t}
\end{equation}
where the two signs originate from the sign of $E/m_b$ and
thus correspond to the baryon and the anti-baryon, respectively.

A static and localized spinor configuration sources the Yang-Mills field via
\begin{equation}
\bar{\cal B}\gamma^{mn}F_{mn}{\cal B}=
\pm \frac12 F_{jk}^a \epsilon^{jki}\langle \sigma_i\tau^a\rangle_{\cal B}
+F_{5i}^a\langle \sigma_i\tau^a\rangle_{\cal B}
\end{equation}
where $\langle \sigma_i\tau^a\rangle_{\cal B}\equiv 2\left[{\cal
U}^\dagger\sigma_i\tau^a{\cal U}\right]$. In order to generate
self-dual or anti-self-dual long-range fields, the spin index and the
gauge index must be locked. For instance, one choice that gives a
long-range self-dual field is ${\cal U}_{\alpha A}=\frac{i}{2}
\epsilon_{\alpha A}$ in which case $\langle
\sigma_i\tau^a\rangle_{\cal B}=-\delta_i^a$ so that the source term
(with the upper sign) is  $-F_{mn}^a\bar \eta^a_{mn}/2$ with the
anti-self-dual 't Hooft symbol $\bar \eta$ ($m,n=1,2,3,5$ and
$a=1,2,3$).

Now assume that such a source appears in a localized form at the
origin. The gauge field far away from the source obeys, after a
gauge choice and ignoring $w$-dependence of the electric coupling,
\begin{equation}
\nabla^2A_m^a= 2g_5(0)\rho^2_{baryon}\bar\eta^a_{mn} \partial_n
\delta^{(4)}(x)
\end{equation}
whose solution is
\begin{equation}
A^a_m= -\frac{g_5(0)\rho^2_{baryon}}{2\pi^2}\bar\eta^a_{mn}
\partial_{n}\frac{1}{r^2+w^2}\, .
\end{equation}
The general shape of the long-range field is consistent with the
identification of the baryon as the instanton. In order to fix
$g_5(0)$, we need to match the states in ${\cal B}$ with {\it
quantized} instanton. This means that the long range field of the
instanton should be modified due to quantum fluctuation of the
instanton along different global gauge directions. How to implement
this quantum effect is explained in detail in \cite{longer}. Here we
briefly sketch the reasoning.

Representing the global gauge rotation in (\ref{magnetic}) as
\begin{equation}
S^\dagger A^a_M  (\tau^a/2)S = A^a_M (\tau^b/2)\;\left(
\tr\left[S^\dagger (\tau^a/2)S\tau^b \right]\right)\,,
\end{equation}
the quantization replaces the quantity in the parenthesis by its
expectation value. Following a reasoning mathematically identical to
that used by Adkins et al~\cite{ANW} for the Skyrme model, we obtain
\begin{equation}
\left\langle\tr\left[S^\dagger (\tau^a/2)S\tau^b
\right]\right\rangle_{\cal B} =-\frac{1}{3}\langle
\sigma_b\tau^a\rangle_{\cal B}=\frac13\delta_a^b\, .
\end{equation}
This allows us to arrive at
\begin{equation}
g_5(0)={2\pi^2}/{3}\, .
\end{equation}
We have ignored $w$-dependence of $g_5(w)$ and $e(w)$. This we
believe is harmless for the very small-size baryon/instanton for the
usual reason.


{\it Four-dimensional effective action of baryons}:---
 After identifying the
relevant five-dimensional action, we perform the KK expansion for
the five-dimensional bulk fields along $w$ to derive a four-dimensional Lagrangian.
The four-dimensional
nucleon arises as the lowest eigenmode of the five-dimensional bulk baryon along
the $w$ coordinate, which should be a mode localized near $w=0$. We
mode-expand ${\cal B}_{L,R}(x^\mu,w)=B_{L,R}(x^\mu)f_{L,R}(w)$,
where $\gamma^5 B_{L,R}=\pm B_{L,R}$ are four-dimensional chiral components, with
the profile functions $f_{L,R}(w)$ satisfying
\begin{eqnarray}
\partial_w f_L(w)+m_b(w) f_L(w) &=& m_B f_R(w)\,,\nonumber\\
-\partial_w f_R(w)+m_b(w) f_R(w) &=& m_B f_L(w)\,.
\end{eqnarray}
The four-dimensional Dirac field for the baryon is then reconstructed as
$B=\left(B_L, B_R\right)^T$. See Ref.\cite{Hong:2006ta} for a
similar model.

We will use the mode expansion in Eq.~(\ref{mode}) to obtain
the baryon couplings to mesons. The eigenmode analysis done in~\cite{Sakai:2004cn}
shows that $\psi_{(2k+1)}(w)$ is even and $\psi_{(2k)}(w)$ is odd
under $w\to-w$, corresponding to vector and axial-vector mesons
respectively. Inserting this expansion into the action, we obtain an
effective Lagrangian density for four-dimensional baryons, 
\begin{equation}
{\cal L}_4 = -i\bar B (\gamma^\mu\partial_\mu +m_B) B+{\cal L}_{\rm
vector} +{\cal L}_{\rm axial}+\cdots\quad,
\end{equation}
where the coupling to vector mesons $ a_\mu^{(2k+1)}$ is given by
\begin{equation}
{\cal L}_{\rm vector}=-i\bar B \gamma^\mu \beta_\mu
 B-\sum_{k\ge 0}g_{V}^{(k)} \bar B \gamma^\mu  a_\mu^{(2k+1)} B\quad,
\end{equation}
and the baryon couplings to axial mesons, including pions, as
\begin{equation}
{\cal L}_{\rm axial}=-\frac{i g_A}{2}\bar B  \gamma^\mu\gamma^5
\alpha_\mu B -\sum_{k\ge 1} g_A^{(k)} \bar B \gamma^\mu\gamma^5
a_\mu^{(2k)} B\quad,
\end{equation}
where various couplings constants $g_{V,A}^{(k)}$ as well as the
pion-nucleon axial coupling $g_A$ are calculated by the overlap of 
wavefunctions. 

For the couplings we have two contributions, one from the minimal
coupling, denoted $g_{A_{min}}$, which is calculable for a given
$\lambda N_c$~\cite{longer} and the other from the magnetic term
(\ref{magnetic}). However, in the large $N_c$ limit, the
contribution from the magnetic term is dominant for all axial
couplings. For the leading axial coupling with the pion,
in particular, the main contribution to $g_A$ becomes 
\begin{equation}
g_A\simeq 0.18N_c  \int dw
\left[\left(2U(w)g_5(w)\over g_5(0)U_{KK} \right)
|f_{L}|^2 \frac{\partial_w\psi_0}{M_{KK}} \right]\, ,
\end{equation}
where we used the previous estimates to find
\begin{equation}
g_5(0){\rho_{baryon}^2/
e^2(0)}\simeq 0.18N_c/ M_{KK}\,.
\end{equation}
For sufficiently localized $f_{L,R}$, which is guaranteed by large
$\lambda N_c$, the leading axial coupling $g_A$ from the magnetic term
is approximately
\begin{equation}
g_A\simeq 0.18 N_c\times\left(4/ \pi\right)\simeq 0.7
(N_c/3)\,.\label{ga}
\end{equation}
We stress that this is
independent of the 't Hooft coupling $\lambda$ and the KK mass
$M_{KK}$ and consequently of the pion decay constant.

Let us now consider electromagnetic responses of the baryons. The
simplest way to obtain the coupling is to include the
electromagnetic field as a nonnormalizable mode of the gauge fields
on D8 branes,
\begin{equation}
A_\mu(x;w)={\cal A}_\mu(x)+i\alpha_\mu(x)\psi_0(w)+\cdots
\end{equation}
The five-dimensional gauge interaction of such a nonnormalizable mode gives the vertex
$\int d^4x\;{\cal A}_\mu J^\mu $, into which we  embed the
electromagnetic interaction.

Here we are interested in isolating the magnetic moment of the
nucleon from the magnetic vertex we found above. For $N_F=2$, for
example, we find the isovector magnetic moment of the nucleon,
$\Delta \mu\equiv\mu_{p}-\mu_{n}$, to be
\begin{equation}
\left(\frac{4g_5(0)\rho^2_{baryon}}{e(0)^2}\right)\int d^4x\;
\left[{\cal U}^\dagger {\bf B}\cdot \sigma {\cal U} \right]
\end{equation}
where ${\bf B}$ is the $SU(N_F)$ part of the magnetic field
strength, embedded into ${\cal F}\equiv\,d{\cal A}+{\cal A}^2$.
Given the normalization, $\tr \,{\cal U}^\dagger  {\cal U}=1/2$, one
can identify $\tr \,{\cal U}^\dagger \sigma {\cal U}$ as the spin
operator ${\bf S}$ of the baryon. Recall that the minimal coupling
of the Dirac field to a vector field produces a universal magnetic
moment $e_{EM}{\bf S}/m_B$. It is  easy to show that this latter
contribution, smaller by relative factor of $1/N_c^2$, {\it adds} to
the above leading contribution.

For $N_F=2$ the $SU(2)$ part of the electromagnetic charge is given
as $diag(1/2,-1/2)$. The ``anomalous magnetic moment" of the nucleon
in which strong-interaction dynamics is encoded is given by
\begin{equation}
\frac{\Delta\mu_{an}}{e_{EM}}=\left[\frac{2g_5(0)
\rho^2_{baryon}}{e(0)^2}\right]\simeq\frac{0.36N_c}{M_{KK}}\,.
\label{mm}
\end{equation}

{\it Discussions}:--- So far, all the computations were
carried out in the large $\lambda$ and large $N_c$ limit, so direct
comparisons with nature would be difficult to justify rigorously.
Nevertheless there
is a line of reasonings which suggests a particular form of
the $O(1)$ correction for $g_A$ and $\mu_{an}$ in the $1/N_c$ expansion.
We would like to close this letter with a brief description and the
resulting comparison with experimental values.

It is based on the following set of observations: (1) The instanton
baryon in five dimensions we obtain is related to a skyrmion in four
dimensions~\cite{Sakai:2004cn}, and shares the same symmetry
structures; (2) in the large $N_c$ expansion, the
skyrmion description is equivalent to the constituent quark
model~\cite{dashen-jenkins-manohar}; and (3) a simple group
theoretic structure in constituent-quark and skyrmion models of
the spin-flavor operators figuring in both $g_A$ and $\Delta\mu_{an}$
suggests that  $N_c$ could be replaced by $N_c+2$~\cite{N+2}. Although we are
ill-equipped to
verify the
above shift directly, let us assume in comparing with nature that such
a shift of $N_c$ takes into account most, if not all, of
the leading corrections.

To proceed we need to fix the parameters of the model. In doing
this, we will adhere to the strategy adopted in baryon chiral
perturbation theory, namely, the parameters are determined in the
meson sector. We take the parameters $f_\pi$, $M_{KK}$ etc as
given in \cite{Sakai:2004cn}. In particular we will take
$M_{KK}\approx 0.94$ GeV and $\lambda N_c\approx 26$. Including
the subleading correction $g_{A_{min}}\approx 0.15$ for the given
$\lambda N_c$, we obtain $g_A\approx 1.32$ and
$\Delta\mu_{an}/\mu_B\approx 3.6$, where $\mu_B$ is the Bohr-magneton.
These should be compared with the experimental values $g^{exp}_A=1.26$ and
$\Delta\mu_{an}^{exp}/\mu_B=3.7$.

Clearly much more study is needed to compute corrections in a more
rigorous and systematic manner. What is intriguing is that even at
this ``crude" leading order, the chiral lagrangian, derived from the
string/gauge duality, is found to describe baryons remarkably well,
which indicates it certainly captures the correct physics of strong
interactions.

More details as well as implications of the infinite tower of the
vector mesons on the vector dominance structure of baryon
electromagnetic form factors will be reported elsewhere
\cite{longer}.

{\bf Note Added:} After this work was completed and has appeared,
a related paper has appeared \cite{Hata}, with a partial overlap
on the instanton size estimate in early part of our manuscript.

This work was supported in part
(D.K.H.) by KOSEF  Basic
Research Program with the grant  No. R01-2006-000-10912-0,
by the KRF Grants, (M.R.) KRF-2006-209-C00002, (H.U.Y.) KRF-2005-070-C00030,
and (P.Y.) by the KOSEF
through the Quantum Spacetime(CQUeST) Center of Sogang University
with the grant number R11-2005-021.

\end{document}